\documentclass{nature}

\usepackage{amsmath}
\usepackage[pdftex]{graphicx}
\usepackage{dcolumn}  
\usepackage{bm}           
\usepackage{dsfont}
\usepackage{color}
\usepackage{braket}

\newcounter{defcounter}
\usepackage{amsthm}

\title{Controlling Nonadiabatic Transitions Through Engineered Ultrafast Laser Fields at Conical Intersections} 

\author{Xuanchao Zhang$^{1}$, Yang-Cheng Ye$^{1}$, Panpan Zhang$^{1}$, Xiangmei Duan$^{1}$, R. J. Dwayne Miller$^{2}$, Fulu Zheng$^{1}$, Ajay Jha$^{3,4}$, Hong-Guang Duan$^{1}$ } 

\begin{document} 

\maketitle 

\begin{affiliations} 
\item Department of Physics, School of Physical Science and Technology, Ningbo University, Ningbo, 315211, P.R. China 
\item The Department of Chemistry and Physics, University of Toronto, 80 St. George Street, Toronto Canada M 5S 3H6 
\item Rosalind Franklin Institute, Harwell, Oxfordshire OX11 0QX, United Kingdom
\item Department of Pharmacology, University of Oxford, Oxford, OX1 3QT United Kingdom  \\ 
\centerline{\underline{\date{\bf \today}}} 
\end{affiliations} 

\begin{abstract} 

In this paper, we investigate coherent control of nonadiabatic dynamics at a conical intersection (CI) using engineered ultrafast laser pulses. Within a model vibronic system, we tailor pulse chirp and temporal profile and compute the resulting wave-packet population and coherence dynamics using projections along the reaction coordinate. This approach allows us to resolve the detailed evolution of wave-packets as they traverse the degeneracy region with strong nonadiabatic coupling. By systematically varying pulse parameters, we demonstrate that both chirp and pulse duration modulate vibrational coherence and alter branching between competing pathways, leading to controlled changes in quantum yield. Our results elucidate the dynamical mechanisms underlying pulse-shaped control near conical intersections and establish a general framework for manipulating ultrafast nonadiabatic processes. 

\end{abstract} 


The aspiration to control molecular dynamics with light has been central to chemical physics since the birth of the laser \cite{Ref1, Ref2, Ref3, Ref4}. Early proposals in the 1980s suggested that chemical outcomes could be directed by exploiting quantum interference between indistinguishable excitation pathways \cite{Ref5, Ref6, Ref7, Ref8, Ref9}. This idea, later formalized as coherent control, held that by tailoring the spectral phase and amplitude of laser pulses one could bias a reaction toward specific products \cite{Ref9, Ref10, Ref11}. These theoretical foundations revealed that even in the weak-field, one-photon regime, interference between vibrational wave packets prepared by a shaped pulse could alter population transfer between states \cite{Ref12, Ref13, Ref14}.

The experimental realization of such control awaited the development of femtosecond pulse shaping in the 1990s, which enabled programmable modulation of spectral phase and amplitude \cite{Ref15, Ref16, Ref17}. These advances moved the field beyond simple state-selective excitation toward dynamic steering of wave packets. A landmark demonstration came from experiments on retinal photoisomerization in bacteriorhodopsin, where one-photon phase shaping altered the cis-trans yield by nearly 20\% under biologically relevant weak-field conditions \cite{Ref18}. These results suggested that shaped pulses could bias how a vibrational wave packet traverses a CI, which is the key nonadiabatic funnel mediating ultrafast internal conversion. The subsequent studies of rhodopsin in high-intensity region has been reported by Bucksbaum \cite{Ref18a}. Yet later reports questioning such effects sparked vigorous debate on whether weak-field coherent control is viable in condensed or biological environments \cite{Ref19}. Meanwhile, theory continued to refine the conditions for one-photon phase control. It was shown that closed systems should be insensitive to phase-only shaping in the strict one-photon limit \cite{Ref20}, but that open or dissipative systems, where environment-induced dephasing occurs on comparable timescales to wave packet evolution, may indeed display control \cite{Ref21, Ref22}. This theoretical perspective positioned environmental decoherence not as an obstacle, but as an enabler of weak-field control by providing the non-commuting operators needed for phase sensitivity.

Beyond weak fields, strong-field and multiphoton regimes have provided spectacular examples of coherent control. Adaptive feedback algorithms demonstrated robust optimization of multiphoton ionization and bond breaking, with molecules themselves effectively “solving their own Schr{\"o}dinger equation” \cite{Ref20}. In condensed matter, coherent control has evolved from early demonstrations of coupled electron-phonon dynamics to modern manipulation of macroscopic order parameters. In a seminal study, Iwai {\em et al.} used double femtosecond pulses to coherently amplify or suppress charge-lattice oscillations in a photoinduced neutral-to-ionic transition, showing that even strongly correlated electron-phonon systems can be directed by pulse timing \cite{Ref23}. More than a decade later, Wall {\em et al.\ } extended this paradigm by coherently controlling a surface structural phase transition, realizing reversible switching between ordered states and highlighting the universality of pulse-shaping strategies in low-dimensional materials \cite{Ref24}. Perhaps most striking are applications in complex biological systems. In neurons, chirped femtosecond pulses were used to tune two-photon excitation of channelrhodopsin-2, thereby modulating photocurrents in intact brain tissue \cite{Ref25}. In protein photochemistry, pulse chirp was found to alter ultrafast electron transfer (ET) in flavodoxin mutants, stretching ET timescales from $\sim$100 fs to over 300 fs and imprinting coherence onto subsequent back-transfer steps \cite{Ref26}. These studies provide compelling evidence that even in noisy, solvated proteins, tailored pulses can steer functional dynamics.

Despite these advances, CIs remain the ultimate test bed for coherent control. CIs are ubiquitous in photochemistry \cite{Ref27, Ref28, Ref29, Ref30, Ref31}, mediating ultrafast transitions between electronic states in processes as diverse as vision, photosynthesis, and photoprotection \cite{Ref32, Ref33, Ref34}. Wave packet branching at a CI is exquisitely sensitive to vibrational phase and coherence, making it a natural candidate for light-field steering \cite{Ref35, Ref36}. Yet in realistic environments, dephasing and phonon coupling rapidly erode coherence, raising the question: Can pulse shaping truly modulate nonadiabatic dynamics near a CI under weak-field conditions? To address this, we investigate a generic three-state molecular model incorporating both local (tuning) and nonlocal (coupling) phonon modes, each coupled to dissipative baths. This framework captures the essential features of CI-mediated dynamics in open systems: multiple electronic states, vibronic coupling, and environmental decoherence. By systematically varying pulse shape and chirp, we probe how wave-packet coherence and population transfer evolve as the system approaches and traverses a CI. This approach tests whether tailored ultrafast pulses can exert control over CI-mediated dynamics in an open system, and it aims to reveal the mechanisms and limitations of coherent control in complex molecular environments.

\section*{Results and Discussion} 

To study the impact of pulse parameters on the quantum efficiency near the CI, we constructed an effective model with selected parameters for it. The constructed PESs are shown in Fig.\ \ref{fig:Fig1} and the detailed modeling and parameters are described in the Materials and Methods section. The detailed process of plotting PESs were shown in the supporting information (SI). The ground state wave packet was vertically excited by artificially edited pulse to initially excited state, labelled as B, in optically directed state evolution. The excited state wave packet subsequently relaxes to the minimum position point labeled C, through the degenerate point defining the CI. Then, the wave packet relax to the D by two pathways, which has been labeled as two arrows in Fig.\ \ref{fig:Fig1}. Thus, the quantum yield of this system can be eventually evaluated by the population locating in C and D. We can monitor the efficiency of quantum yield by simply defining of $\rm pop(D)/(pop(C)+pop(D))$ after the wave packet reaches equilibrium.   

\subsection{Pulse Chirp Modulation of Wave-packet Dynamics and Quantum Yield} 

With the constructed PESs, we employed the hierarchy equation of motion (HEOM) \cite{HEOM1, HEOM2} approach and calculated the wave packet evolution after exciting by artificially edited pulse. The linear vibronic Hamiltonian and system-bath model have been described in the Materials and Methods section. Moreover, the pulse formula (Gaussian shaped pulse) and the parameters were presented as well. With varying $\eta$ (dimensionless chirp parameter of the Gaussian laser pulse), we were able to modulate the pulse duration and the associated chirp. For this, we have plotted the laser pulses with varying of $\eta$ = 5, 0 and -10. We show the pulse profiles as blue, green and red solid lines (time domain) in Fig.\ \ref{fig:Fig2}(a), (b) and (c), respectively. With the edited pulses, we calculated the time evolved population dynamics and preformed the projection procedure to show the detailed motions of excited-state wave packet along the reaction coordinates (tunning mode $Q_{t}$). The detailed projection procedure are shown in the SI. By this, we have obtained the wave-packet dynamics and plotted in Fig.\ \ref{fig:Fig2}(d) to (f) with $\eta$ = -10, 0 and 5, respectively. In Fig.\ \ref{fig:Fig2}(d), we observed, at early time, the wave packet was mainly located at higher electronic excited state. At roughly t = 200 fs, the population evolves on to the lower adiabatic electronic state. The wave packet mainly becomes located at $Q_{t}$ = -1 at $\sim$200 fs. The wave packet rapidly transitions to two minimum ($Q_{t}$ = -1.8 and 1.7, which has been marked as blue solid lines) within the initial 800 fs. The wave packet clearly oscillates within this time window of 5 cycles, which yields a period of $\sim$110 fs, which corresponds to the frequency of the tuning mode, $\Omega_{t}$ = 300 cm$^{-1}$.  The wave packet reaches an effectively static position at around t = 1.4 ps. At this point, there is clearly a potential barrier between two minima, which makes the clear definition of quantum efficiency by ratio of the shadow areas on both sides (t from 1.8 ps to 2 ps), which has been marked as red dashed boxes. By this consideration, we obtained a quantum yield for the system (with $\eta$ = -10) of 0.436. Moreover, we repeated the same procedure and plotted the wave packet dynamics in Fig.\ \ref{fig:Fig2}(e) with $\eta$ = 0. We observed, at initial time T =  0 fs, the wave packet in mainly locating on the higher excited electronic state. At T = 200 fs, the wave packet passes through the degenerate point and reaches the lower excited state. However, it shows significantly different dynamics compare to the case of $\eta$ = -10 in Fig.\ \ref{fig:Fig2}(d). We show that the wave packet reaches the maximum at shorter timescales due to the short pulse duration with $\eta$ = 0. We have shown the detailed differential motions on short timescales in the SI. The wave packet moves and oscillates to its equilibrium at $\sim$800 fs. The quantum efficiency can be effectively measured at times from 1.8 ps to 2.0 ps, which has been marked as a red dashed box in Fig.\ \ref{fig:Fig2}(e). The calculated quantum efficiency is 0.406 within the shadowed area. Follow the same procedure, we have calculated the wave-packet dynamics of the system with $\eta$ = 5. The resulting data are shown in Fig.\ \ref{fig:Fig2}(f). Compared to the other two cases, the initial wave packet is also mainly located at the electronic higher excited state. They evolve to the lower excited state at T = 200 fs. However, the wave packet shows a dramatic difference compared to the calculated results with $\eta$ = -10 and 0 in Fig.\ \ref{fig:Fig2}(d) and (e). The wave packet stays at $Q_{t}$ = -1.8 for relatively short times compared to the case of Fig.\ \ref{fig:Fig2}(d) and relax with oscillations corresponding to vibrational coherence. However, the observed oscillations do not exhibit a single, well-defined period, indicating that different components of the wave packet evolve out of phase. As a consequence of this incoherent oscillatory behavior, the calculated quantum efficiency reaches its minimum value of 0.396.

We have plotted the laser spectrum (in frequency domain) employed in the calculation in Fig.\ \ref{fig:Fig2}(g). By varying parameter of $\eta$, we were able to globally study the impact of pulse chirp for the quantum efficiency in the passage through the CI. We have plotted the quantum efficiency versus parameter of $\eta$ in Fig.\ \ref{fig:Fig2}(h). It shows a very interesting curve with three distinct features. For instance, the quantum efficiency reaches to 0.406 for $\eta$ = 0 (no chirp) and can not be increased until the chirp reaches magnitude of $\eta$ = $\pm$1.8. Moreover, we also observe a relatively smoother curve at $\eta$ = 5 in Fig.\ \ref{fig:Fig2}(h). For this, we have magnified the wave packet dynamics in a shorter time window. We have shown the resulting data in Fig.\ \ref{fig:Fig3}. In Fig.\ \ref{fig:Fig3}(a), we presented the dynamics with $\eta$ = -10. With this chirp, we know that the lower frequency components of the pulse excite the wavepacket  first, followed by the high fequency terms, which can be directly observed in the pulse profile in time domain (Fig.\ \ref{fig:Fig2}(a)). Thus, the wave packet arrives at the lower excited electronic state with a broader and smoother temporal profile, primarily due to the longer duration of the negatively chirped pulse and the reduced kinetic energy arising from the stronger weighting of excitation components in the lower-frequency range. We further observe that the wave packet remains within the time window of [150, 250 fs] before subsequently relaxing, indicating that it initially lacks sufficient momentum to relax to the lower regions of the PESs. The detailed schematic diagram of wave-packet motion is shown in Fig.\ \ref{fig:Fig3}(b).

In contrast, the wave packet motions show different dynamics with short pulse excitation, $\eta$ = 0, which is shown in Fig.\ \ref{fig:Fig3}(c). It reaches a maxima at T = 200 fs and it stays for a relatively short time. The wave packet relaxes to the lower part of PESs immediately after the excitation. This means that the short, unchirped pulse excites the wave packet to the higher excited state with its full kinetic energy. The wave packet passes through the degenerate point and reaches to the lower excited state. Due to the short interaction time, the wave packet involves higher and lower vibrational levels, leading to propagation to the lower electronic excited state almost simultaneously. This allows the clear observation of vibrational coherence since the wave packet is constructed with a perfect synchronization of the various vibration components. In the case of Fig.\ \ref{fig:Fig3}(c), we show the detailed motions of the wave packet with short time detection window. The schematic motions of the wave packet with short pulse duration has been shown in Fig.\ \ref{fig:Fig3}(d). We also show that with parameter $\eta$ = 5, the higher-frequency part reaches and excites the sample before the low-frequency part. Thus, wave packet reaching to the higher excited state and relaxes to the lower one. Due to the imperfect of chirp parameter, the wave packet on lower and higher vibrational levels can not relax within the same step, thus, the relaxation induces a phase mismatch in the different frequency components of the wave packet upon reaching the lower electronic state. Thus, we observed a relatively noisy wave packet in the time window of 150 fs to 250 fs in Fig.\ \ref{fig:Fig3}(e). To understand it better, we have shown the schematic diagram with vectors of wave packet motion in Fig.\ \ref{fig:Fig3}(f).

\subsection{Effect of Vibrational Energy Gap on Control Outcomes} 

In order to study the quantum efficiency with different configurations of PESs, we have modulated the parameters and constructed the PESs with different energy gaps, as shown in Fig.\ \ref{fig:Fig4}. We plotted the PESs with an energy gap of $\Delta\epsilon$ = 1400 cm$^{-1}$ in Fig.\ \ref{fig:Fig4}(a). We performed the calculations and showed the quantum efficiency with varying of $\eta$ from -10 to 10 in Fig.\ \ref{fig:Fig4}(b). We observed the quantum efficiency changed from 0.156 to 0.13. Moreover, we also observed the derivatives of the curves changed at around $\eta$ = 0 and 5. In this case, we determined that the efficiency can be improved by 3\%. In addition, we show the PESs with energy gap $\Delta\epsilon$ = 600 cm$^{-1}$. In this case, we find that quantum efficiency can be changed from 0.725 to 0.705. These results are shown in Fig.\ \ref{fig:Fig4}(c) and (d). The results with energy gap $\Delta\epsilon$ = 200 cm$^{-1}$ and the resulting quantum efficiency are shown in (e) and (f). Interestingly, we also observed the change in slope for the quantum efficiency curve for $\eta$ = 0 and 5. This finding suggests that the resonant vibrational period of the wavepacket plays a role in how chirp affects the outcome: chirps that synchronize or desynchronize with the wavepacket’s natural oscillation period produce noticeable kinks in the yield trend. We also analysed wavepacket dynamics for alternative potential-energy-surface configurations (see SI), confirming that the quantum efficiency remains tunable via the chirp parameter $\eta$ and depends on the energy separation of product minima and vibrational level spacing. We further examined geometric-phase effects during passage through the CI. Although chirped pulses suppress wavepacket cancellation, we find no evidence that the geometric phase directly modulates the quantum efficiency; the suppression instead originates from phase instability during photoexcitation.

\subsection{Influence of Vibrational Coherence Lifetime (System-Bath Coupling)} 

We further examined the quantum efficiency versus the coherence lifetime of vibronic modes. For this, we varied the reorganization energy and control the lifetime of vibrational coherence. We then repeated the calculations of wave-packet dynamics and also examined the quantum efficiency with changes of system-bath couplings. Based on the calculations, we have obtained the population and resulting wave-packet dynamics presented in Fig.\ \ref{fig:Fig5}. We plotted the motions of wave packet with $\lambda$ = 5, 8 and 20 cm$^{-1}$ from Fig.\ \ref{fig:Fig5}(a) to (i), which shows the coherence with relatively long, intermediate, and short timescales. It can be directly observed in the oscillatory dynamics along the reaction coordinate. We then estimated the quantum efficiency with long population times. The quantum yields versus the magnitude of reorganization energy $\lambda$ and chirp parameter $\eta$ are shown in Fig.\ \ref{fig:Fig5}(j). The general finding is that longer vibrational coherence times markedly improve the degree of control attainable. For the smallest $\lambda$ (weak system-bath coupling), the yield curve spans a larger range, indicating up to $\sim$5-6\% absolute enhancement of yield with optimal chirp. In contrast, for strong bath coupling ($\lambda$ = 20 cm$^{-1}$, fast dephasing), the yield is nearly insensitive to chirp, with $<$1\% variation. This makes physical sense: if vibrational coherence is lost quickly, the wavepacket’s phase information is erased before the pulse can direct it, nullifying the benefit of coherent shaping. Conversely, if coherence persists, the shaped pulse can effectively leverage interference effects to influence the branching. We also confirm that negative chirp gives the highest yield in all cases (e.g. at $\lambda$ = 5 cm$^{-1}$, the best yield comes from red-first pulses), consistent with our earlier results. These observations highlight that vibrational coherence is the central resource enabling control: extending the coherence lifetime (either by reducing environmental dissipation or by other means) directly translates to a larger controllable yield window. In our model, the strong nonadiabatic coupling at the CI creates vibrational coherences that are not merely epiphenomena but actively mediate the reaction pathway.

\section*{Discussions}

Our results demonstrate that engineered ultrafast pulses can influence nonadiabatic transitions at conical intersections, although the achievable control is intrinsically limited. By shaping the spectral phase of the excitation pulse, we obtain modest variations of approximately $\sim$5 - 10\% in the product yield of an isomerization-like model system, consistent with the mixed outcomes reported in one-photon coherent control studies. In principle, pulse shaping can manipulate vibrational wavepackets and bias their passage through a CI. In realistic molecular environments, however, rapid dephasing, multi-mode nuclear motion, and strong system-bath coupling tend to suppress the delicate wavepacket interference required for robust control under weak-field conditions. Our simulations place these observations on a unified footing. Using an idealised two-state CI model, we clearly resolve a dependence of product yield on pulse shape, but the magnitude of the effect remains small, reflecting the intrinsically subtle nature of phase-only control in the one-photon limit. While the model preserves vibrational coherence and therefore represents an upper bound on controllability within this framework, it captures the essential mechanism: nonadiabatic coupling at the CI generates a coherent superposition of reaction pathways, and the relative phase between these components governs the branching outcome. Control is thus fundamentally constrained by the lifetime of vibrational coherence. Within this model, artificially shaped pulses can enhance or suppress the quantum efficiency by more than $\sim$5\%. However, the simplicity of the two-state, two-mode description allows the nonadiabatic dynamics to be interpreted transparently and may underestimate the potential for control in more realistic systems. In complex molecular environments, such as rhodopsin, additional structural degrees of freedom, most notably torsional motion about the C=C bond, can couple strongly to photoexcitation and may offer enhanced control under suitably tailored pulses, as also suggested in earlier studies\cite{Valentyn 2006}. At the same time, recent weak-field experiments highlight the sensitivity of such effects to decoherence\cite{Ref19}. Recent isotope-resolved studies on rhodopsin further suggest that vibrational phase at the CI can influence quantum yield\cite{Mathies_isotope}. In this context, our work provides a clear dynamical framework for understanding how pulse properties shape wavepacket motion near a CI and how this, in turn, modulates quantum efficiency.

A key insight is that vibrational coherence and nonadiabatic coupling act cooperatively. Negatively chirped (red-first) pulses gently launch the wavepacket and better preserve phase synchrony along the reaction coordinate, thereby favouring one product channel. In contrast, positively chirped or poorly timed pulses induce dephasing and diminish interference. This asymmetry reflects a general principle of coherent control, whereby down-chirped excitation deposits energy more adiabatically and enhances slower pathways, whereas up-chirped excitation preferentially activates competing motions. At the same time, our results highlight the limitations of single-pulse control near CIs. In complex polyatomic systems, multiple coupled vibrational modes, additional decay channels, and environmental dissipation further accelerate dephasing and dilute control. Consequently, achieving substantial yield modulation will likely require strategies beyond single-pulse phase shaping. Multi-pulse or pump-repump schemes, which exploit recurrent wavepacket motion and selectively reinforce coherence, offer a promising route to stronger control. More recently, multi-pulse configuration has been developed to modulate the quantum efficiency. For instance, Ishikawa and Ropers works \cite{Shinya Koshihara coherent control, Penpeng's work, Ropers Nature CDW LEED}. The quantum efficiency can be significantly enhanced by two pulses with carefully design with resolved key vibrational modes. The uncovered modes during the phase transition can be revealed and the duration between two pulses can be achieved by the time delay. The enhancement of efficiency when the second pulse reaching with perfect matching of period. This control strategy can be easily understood by our work with periodic oscillations of the wave packet. Overall, our work establishes that coherent control at CIs is physically real but fundamentally constrained. The extent to which it can be realised in complex molecular systems is governed by the balance between ultrafast vibronic dynamics and environmental decoherence, providing clear guidelines for the design of future control strategies.

\section*{Conclusion} 

Our work underscores how ultrafast pulse shaping can influence photochemical outcomes by harnessing vibrational coherence at CIs. We have provided a detailed analysis linking pulse characteristics to wavepacket dynamics and reaction yields. The broader implication is that even in seemingly ultrafast and irreversible processes like internal conversion, there exist windows of opportunity (on the order of tens to hundreds of femtoseconds) where the course of the reaction is not predetermined but can be nudged by coherent light. Future studies on more complex systems will help determine how far this principle can be generalized. While single-pulse phase modulation yields incremental control, combining it with multi-pulse sequences or advanced feedback learning algorithms may achieve more dramatic steering of chemical reactions. Ultimately, the interplay of pulse shaping, vibrational dynamics, and nonadiabatic couplings revealed here provides a foundation for designing quantum light-induced control strategies in photochemistry, from simple models to biological and material systems. By targeting the right mode at the right time, whether through chirp, pulse timing, or spectral tailoring, it becomes feasible to tilt reaction pathways in a desired direction, offering a tantalising degree of control over molecules in motion.

%


\section*{Materials and Methods}
\subsection{Modeling and parameters} 

\begin{figure}[h!]
\begin{center}
\includegraphics[width=11.0cm]{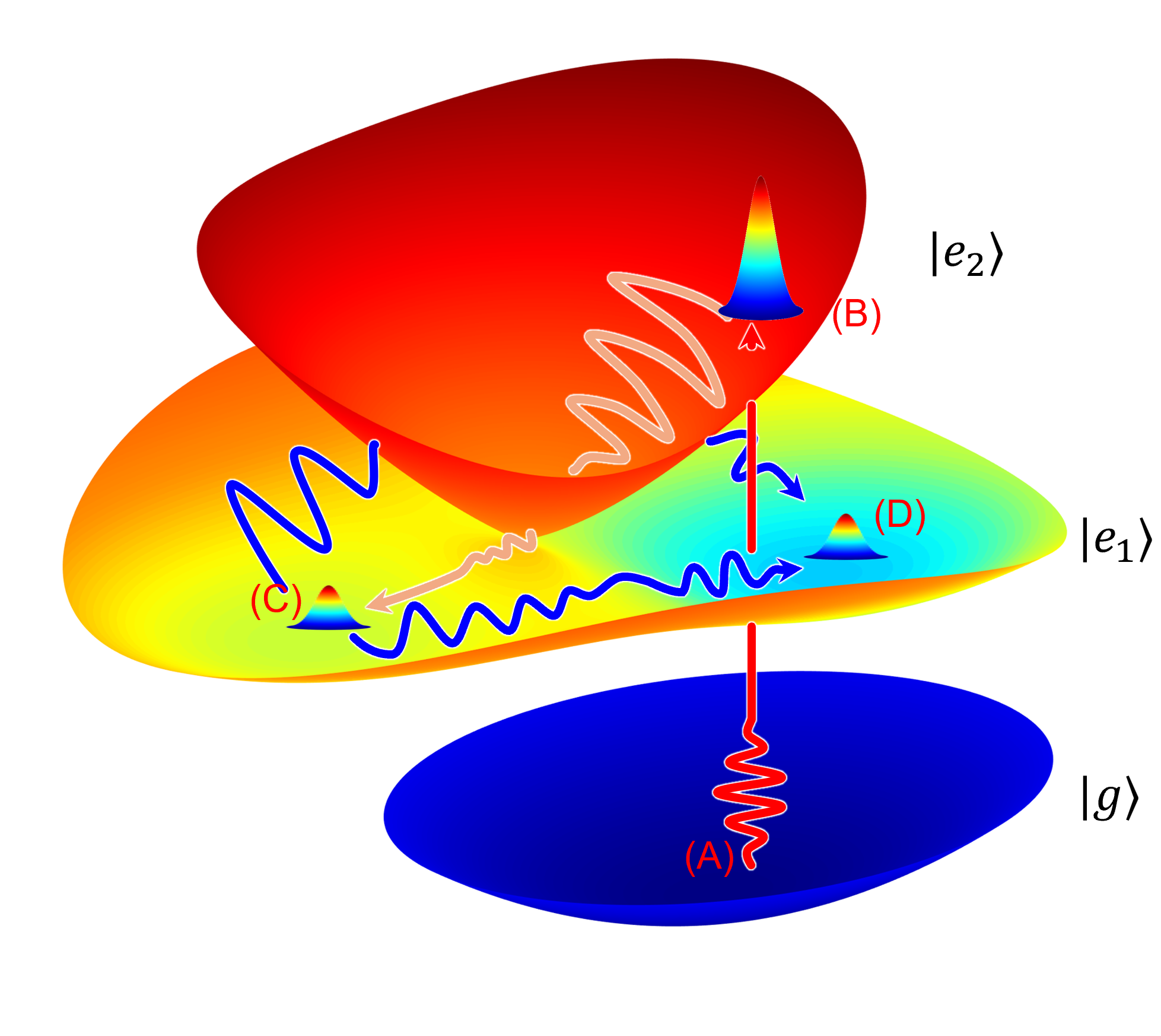} 
\caption{\label{fig:Fig1} Potential energy surfaces (PESs) and the CI between two electronically excited states. The vertical excitation move wave packet from A to B. The wave packet relax to C and further to D after delay time. }  
\end{center}
\end{figure}

\begin{figure}[h!]
\begin{center}
\includegraphics[width=15.0cm]{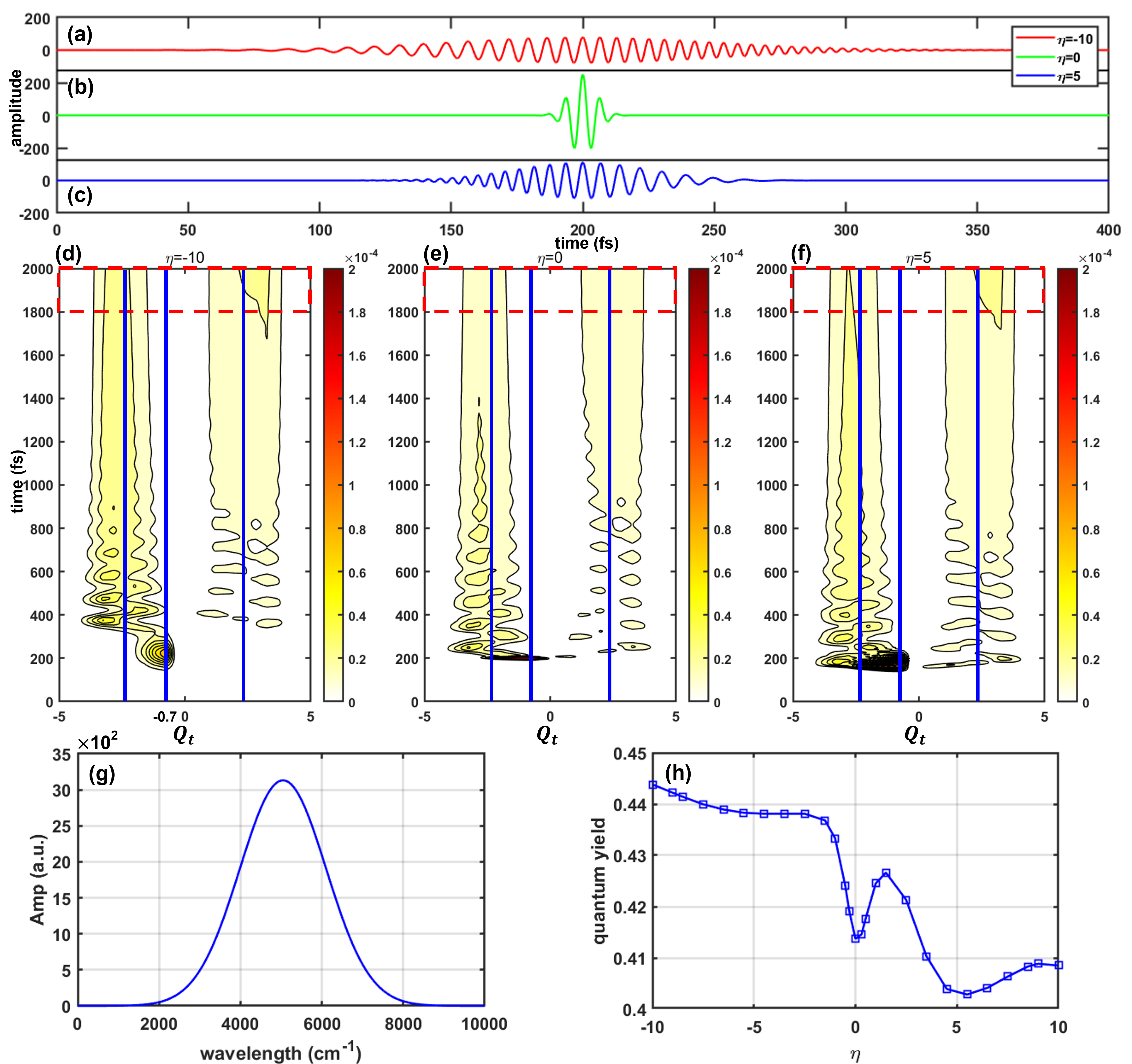} 
\caption{\label{fig:Fig2} (a-c) pulse information in time domain with different $\eta$. Wave-packet dynamics of the population transfer in the CI with different parameters of $\eta$ in (d), (e) and (f). (g) the pulse profile in the frequency domain. (h) the dependence of quantum yield with different $\eta$. } 
\end{center}
\end{figure}

\begin{figure}[h!]
\begin{center}
\includegraphics[width=14.0cm]{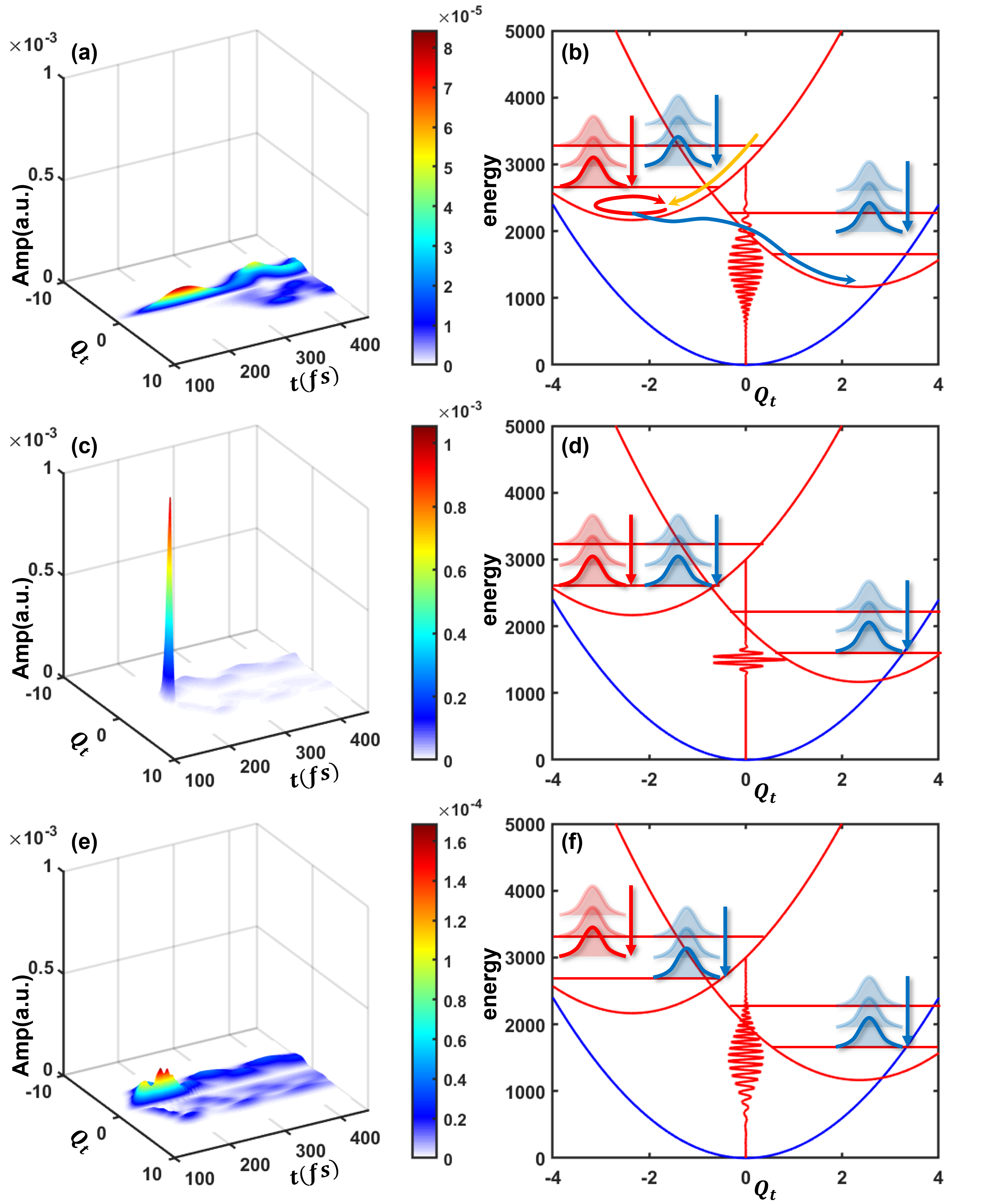} 
\caption{\label{fig:Fig3} Wave-packet dynamics evolving with time are shown in (a), (c) and (e). The detailed motions of the wave packets are shown in (b), (d) and (f), respectively. } 
\end{center}
\end{figure}

\begin{figure}[h!]
\begin{center}
\includegraphics[width=14.0cm]{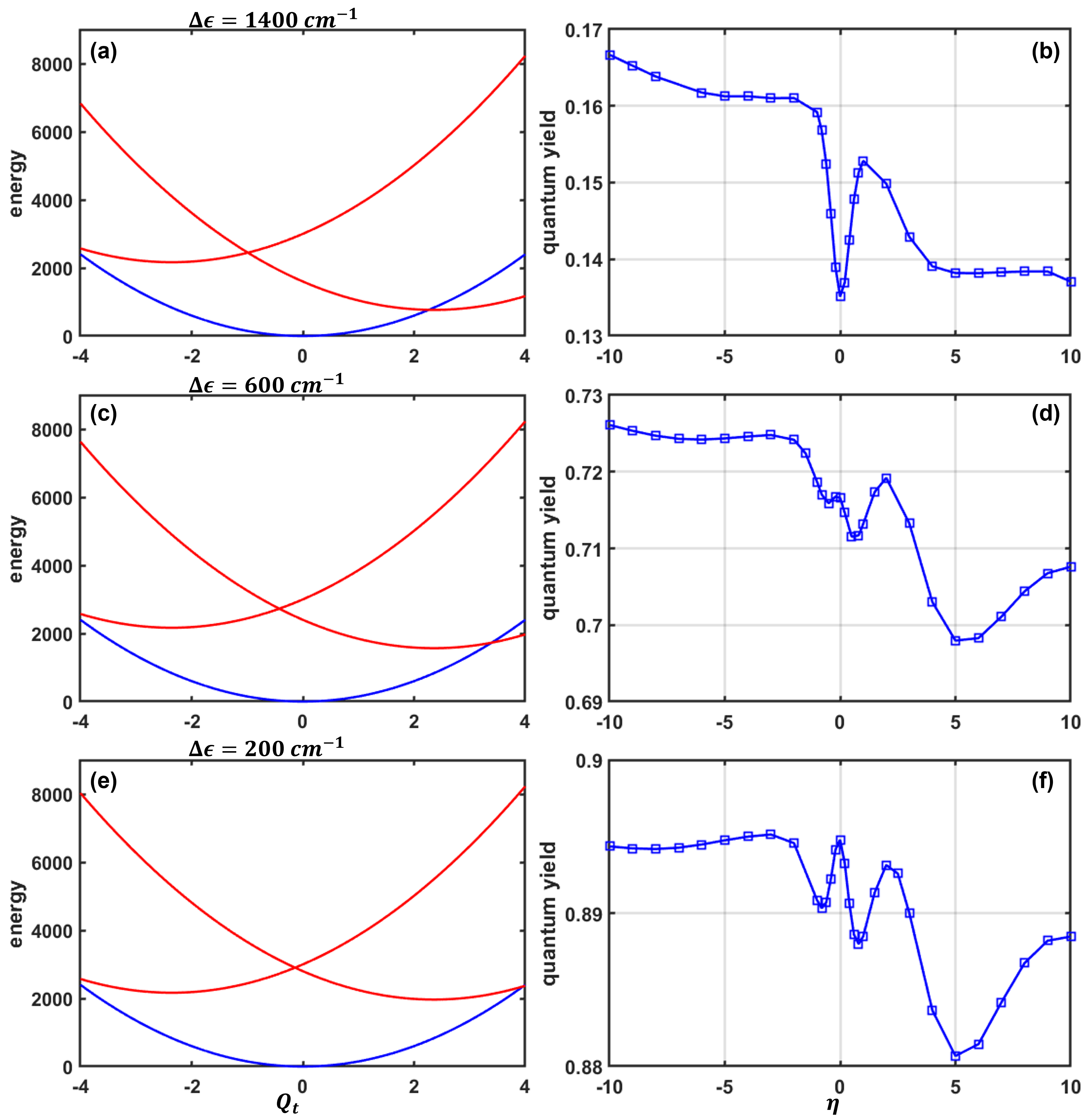} 
\caption{\label{fig:Fig4} The modulation of parameters to yield the PESs with different energy gaps in (a), (c) and (e). The associated quantum efficiency with varying of $\eta$ are shown in (b), (d) and (f), respectively. } 
\end{center}
\end{figure}

\begin{figure}[h!]
\begin{center}
\includegraphics[width=14.0cm]{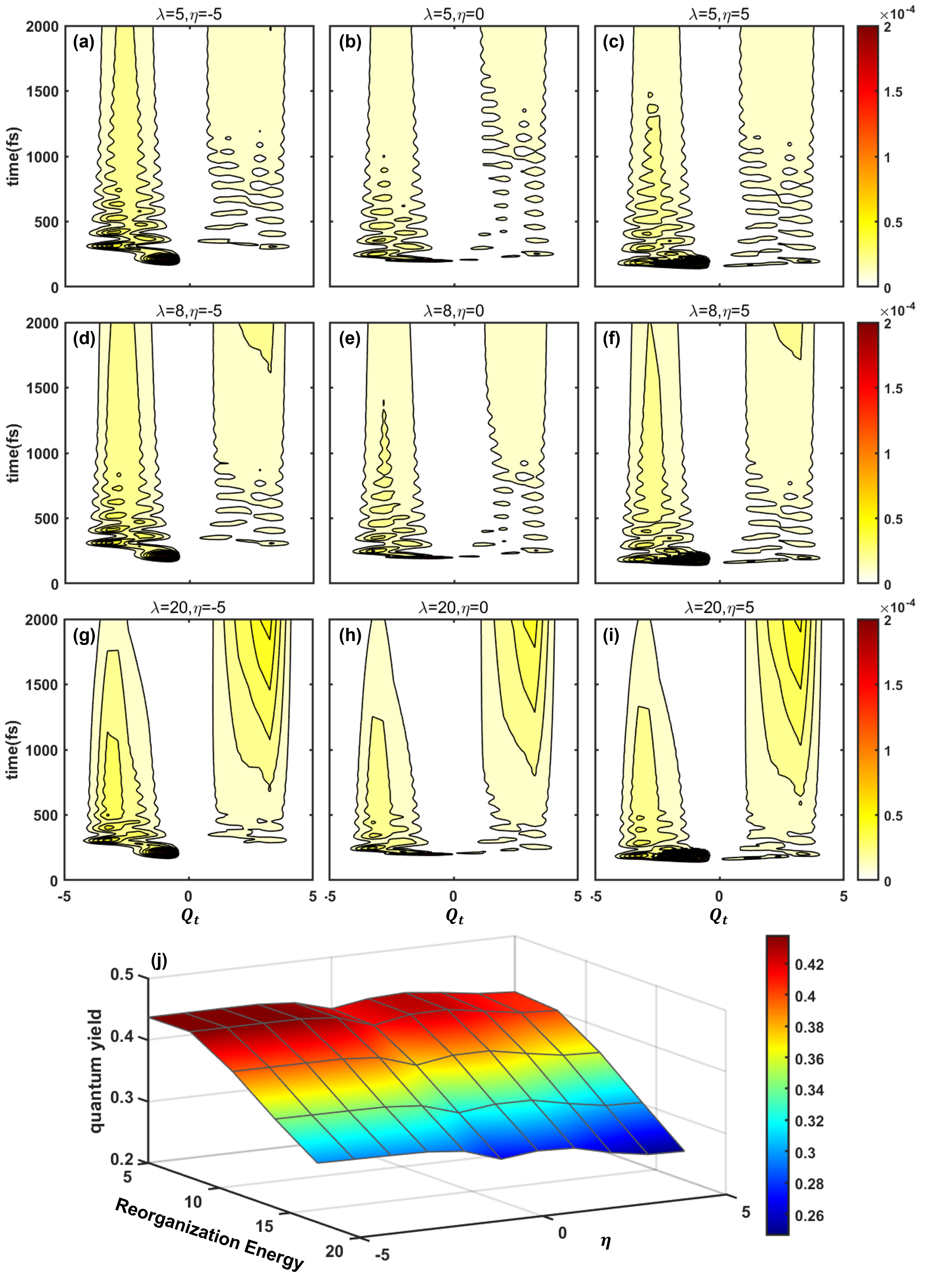} 
\caption{\label{fig:Fig5}The wave-packet dynamics with varying of reorganization energies $\lambda$ and chirp parameters $\eta$. The wave-packet dynamics with selected parameters are shown from (a) to (i). The quantum yields versus $\eta$ and $\lambda$ are shown in (j). } 
\end{center}
\end{figure}

We consider a molecular system with three electronic states to study the coherent dynamics controlling in the CI. In this model, each of electronic state is coupled to both an intramolecular local phonon (tuning mode) and an intermolecular nonlocal phonon (coupling mode). These two phonons are additionally coupled to their dissipative environment, respectively. The Hamiltonian includes molecular and molecular-environment coupling parts. The molecular Hamiltonian can be written as  
\begin{equation}
\begin{aligned}
\label{eq:sysham} 
H_{\rm mol}& = \ket{g}h_{g}\bra{g} + \ket{e_{1}}h_{e1}\bra{e_{1}} + \ket{e_{2}}h_{e2}\bra{e_{2}}\\& + (\ket{e_{1}}V\bra{e_{2}} + h.c.), 
\end{aligned}
\end{equation}
with $h_{e1}=\epsilon_{1} + h_{g} + \kappa_{1}Q_{t}$, $h_{e2}=\epsilon_{2} + h_{g} + \kappa_{2}Q_{t}$ and the electronic coupling strength $V = V_{0} + \Lambda Q_{c}$, where $h_{g}=\frac{1}{2}\sum_{i=t,c}{\frac{P_{i}^{2}}{2\mu} + \frac{1}{2}\mu\Omega_{i}^{2}Q_{i}^{2}}$. $P_{t/c}$ and $Q_{t/c}$ denotes the momentum and coordinate operator of the mode with the vibrational frequency $\Omega_{t/c}$, $\mu$ is the mass of the particle. The local phonon is coupled to the electronic state, the latter having the site energy $\epsilon_{1/2}$, via the vibronic coupling  strength $\kappa_{1/2}$. Both excited states are electronically coupled via a Coulomb interaction with a static part $V_{0}$ and a dynamic Peierls part due to their coupling to the nonlocal phonon with strength $\Lambda$. In addition, to include the dissipation, we assume that the two effective modes are coupled to the displacements of harmonic bath oscillators, thus, we have 
\[
\label{eq:bathham} 
H_{\rm env} = \sum_{i=\rm 1,2}\sum_{\alpha}\left[\frac{p^{2}_{i,\alpha}}{2m_{i,\alpha}} +\frac{m_{i,\alpha}\omega^{2}_{i,\alpha}}{2} \left( x_{i,\alpha}+\frac{c_{i,\alpha}Q_{i}}{m_{i,\alpha} \omega^{2}_{i,\alpha}}  \right)^{2} \right]\, .
\]
Here, $\alpha$ is the index of the bath modes. $p_{i,\alpha}$, $m_{i,\alpha}$ and $\omega_{i,\alpha}$ are the momenta, mass and frequency of the $\alpha$-th mode coupled to $Q_{i}$, and the $c_{i,\alpha}$ is the coupling coefficient. The baths are characterized by the spectral  densities $J_{t/c}(\omega)=2\lambda\frac{\omega\gamma}{\omega^{2}+\gamma^{2}}$. For simplicity, we assume two Drude spectral densities throughout this work, in this function, $\lambda$ is the reorganization energy and $\gamma$ is the cutoff frequency. For the vibronic interaction, two excited potential energy surfaces are shifted relative to the ground state, which is characterized by the tuning parameters $\Delta_{1}$ and $\Delta_{2}$. We have $\kappa_{1}=\Delta_{1}\Omega_{t}$/$\sqrt{2}$ and $\kappa_{2}=\Delta_{2}\Omega_{t}$/$\sqrt{2}$. To give a concrete example, we assign $\Omega_{t}$ = 300 cm$^{-1}$. We assume $\Delta_{2}=-\Delta_{1}=2.357$ to modulate a shift between first excited states and second excited states, so that the degenerate  point will locate at q = -1 position. In addition, the coupling strength of the two modes is set to $\Lambda$ = 200 cm$^{-1}$. The potential energy surface is shown in Fig.\ \ref{fig:Fig1}(a). We perform the calculations at room temperature, T = 300 K. 

To considering the light-matter interaction, the time-dependent Hamiltonian can be written as 
\begin{equation}
\label{eq:sysham} 
H_{\rm tot} = H-H_{F}(t),
\end{equation}
the $H_{F}(t)$ is the light-matter interaction part. In the dipole approximation and the rotating-wave approximation (RWA), the Hamiltonian $H_{F}(t)$ is given by 
\begin{equation}
\label{eq:sysham} 
H_{F}(t) = -\hat{\mu}_{g,e2} \cdot E(t),
\end{equation}
the $\hat{\mu}_{g,e2}$ denotes the transition dipole moment and $E(t)$ the Gaussian shaped pulse, the latter can be written as 
\begin{equation}
\label{eq:sysham} 
E(\eta,t) = \frac{1}{2} E_{max}(\eta) F(\eta,t) e^{i\Psi(\eta,t)}e_{z} + c.c.
\end{equation}
the $E_{max}$ represents the maximum pulse amplitude, $F(\eta,t)$ represents the temporal envelope of a chirped pulse and $\eta$ represents dimensionless chirp parameter of the Gaussian laser pulse. It controls both the frequency chirp (instantaneous phase) and the effective pulse duration, thereby shaping how the wave packet is excited and evolves near the CI. $\Psi(\eta,t)$ is the instantaneous phase, respectively, which is given by
\begin{equation}
\label{eq:sysham} 
E_{max}(\eta) = \frac{E_{0}}{(1+\eta^{2})^{\frac{1}{4}}},
\end{equation}
\begin{equation}
\label{eq:sysham} 
F(\eta,t) = exp(\frac{-t^{2}}{2T(\eta)^{2}}),
\end{equation}
\begin{equation}
\label{eq:sysham} 
\Psi(\eta,t) = \omega_{0}t - \frac{\eta t^{2}}{2T_{0}^{2}(1+\eta^{2})}.
\end{equation}
Here, $T{\eta}$ represents the effective pulse duration, which are given by
\begin{equation}
\label{eq:sysham} 
T(\eta) = T_{0}\sqrt{1+\eta^{2}}.
\end{equation}

%
\begin{addendum}
\item This work was supported by the National Key Research Development Program of China (Grant No.\ 2024YFA1409800), NSFC Grants with No.\ 12274247, 12404310, 12504279 and W2421021 , Yongjiang talents program with No.\ 2022A-094-G and 2023A-158-G, Ningbo International Science and Technology Cooperation with No.\ 2023H009, ``Lixue+'' Innovation Leading Project and the foundation of national excellent young scientist. The Next Generation Chemistry theme at the Rosalind Franklin Institute is supported by the EPSRC (V011359/1 (P)) (A.J.). 

\item[Supporting information] The details of projection calculations, the HEOM quantum master equation and the quantum efficiency with different configurations of PESs. The evidence of geometric phase and the calculations with different pulse shapes and durations are also included in the supporting information.

\item[Competing Interests] The authors declare that they have no competing financial interests. 

\item[Author contributions] H.G.D. conceived the research concept and discussed with A.J.; X.Z. performed the calculations and discussed with Y. Y.; H.G.D. and A.J. wrote the initial draft and all authors wrote the final manuscript. 

\item[Correspondence] Correspondence of paper should be addressed to F.Z. ~(zhengfulu@nbu.edu.cn), A.J. ~(Ajay.Jha@rfi.ac.uk) and H.-G.D. ~(email: duanhongguang@nbu.edu.cn). 

\end{addendum}
%
\newpage

\newpage

\newpage

\newpage

\newpage


\begin{thebibliography}{11} 

\bibitem{Ref1} Zinth, W., Laubereau, A. \& Kaiser, W. The long journey to the laser and its rapid development after 1960. The European Physical Journal H 36, 153-181 (2011). https://doi.org/10.1140/epjh/e2011-20016-0

\bibitem{Ref2} Brumer, P. \& Shapiro, M. Laser control of molecular processes. Annual review of physical chemistry 43, 257-282 (1992).
 
\bibitem{Ref3} Gordon, R. J. \& Rice, S. A. Active control of the dynamics of atoms and molecules. Annual review of physical chemistry 48, 601-641 (1997). 

\bibitem{Ref4} Bergmann, K., Theuer, H. \& Shore, B. Coherent population transfer among quantum states of atoms and molecules. Reviews of Modern Physics 70, 1003 (1998). 

\bibitem{Ref5} Laforge, F. O., Lee, J. \& Rabitz, H. A. The Early Era of Laser-Selective Chemistry 1960-1985: Roots of Modern Quantum Control. The Journal of Physical Chemistry Letters 14, 5283-5296 (2023). https://doi.org/10.1021/acs.jpclett.3c00678

\bibitem{Ref6} Shapiro, M. \& Brumer, P. Laser control of product quantum state populations in unimolecular reactions. The Journal of Chemical Physics 84, 4103-4104 (1986). https://doi.org/10.1063/1.450074

\bibitem{Ref7} Brumer, P. \& Shapiro, M. Control of unimolecular reactions using coherent light. Chemical physics letters 126, 541-546 (1986). 

\bibitem{Ref8} Tannor, D. J. \& Rice, S. A. Control of selectivity of chemical reaction via control of wave packet evolution. Journal of Chemical Physics 83, 5013-5018 (1985). 

\bibitem{Ref9} Brumer, P. \& Shapiro, M. One photon mode selective control of reactions by rapid or shaped laser pulses: An emperor without clothes? Chemical physics 139, 221-228 (1989). 

\bibitem{Ref10} Katz, G., Ratner, M. A. \& Kosloff, R. Control by decoherence: weak field control of an excited state objective. New Journal of Physics 12, 015003 (2010). 

\bibitem{Ref11} Spanner, M., Arango, C. A. \& Brumer, P. Communication: Conditions for one-photon coherent phase control in isolated and open quantum systems. The Journal of chemical physics 133 (2010). 

\bibitem{Ref12} Warren, W. S., Rabitz, H. \& Dahleh, M. Coherent control of quantum dynamics: the dream is alive. Science 259, 1581-1589 (1993). 

\bibitem{Ref13} Rabitz, H., de Vivie-Riedle, R., Motzkus, M. \& Kompa, K. Whither the future of controlling quantum phenomena? Science 288, 824-828 (2000). 

\bibitem{Ref14} Shapiro, M. \& Brumer, P. Principles of the quantum control of molecular processes.  (2003).

\bibitem{Ref15} Herek, J. L., Materny, A. \& Zewail, A. H. Femtosecond control of an elementary unimolecular reaction from the transition-state region. Chemical Physics Letters 228, 15-25 (1994). https://doi.org/https://doi.org/10.1016/0009-2614(94)00910-4

\bibitem{Ref16} Sheehy, B., Walker, B. \& DiMauro, L. F. Phase Control in the Two-Color Photodissociation of ${\mathrm{HD}}^{+}$. Physical Review Letters 74, 4799-4802 (1995). https://doi.org/10.1103/PhysRevLett.74.4799

\bibitem{Ref17} Weinacht, T. C., Ahn, J. \& Bucksbaum, P. H. Controlling the shape of a quantum wavefunction. Nature 397, 233-235 (1999). https://doi.org/10.1038/16654

\bibitem{Ref18} Prokhorenko, V. I. {\em et al.} Coherent control of retinal isomerization in bacteriorhodopsin. Science 313, 1257-1261 (2006). 

\bibitem{Ref18a} A. C. Florean, D. Cardoza, J. L. White, J. K. Lanyi, R. J. Sension, P. H. Bucksbaum, Control of retinal isomerization in bacteriorhodopsin in the high-intensity regime. Proc. Natl. Acad. Sci. U.S.A. 106, 10896-10900 (2009).

\bibitem{Ref19} Liebel, M. \& Kukura, P. Lack of evidence for phase-only control of retinal photoisomerization in the strict one-photon limit. Nature Chemistry 9, 45-49 (2017). https://doi.org/10.1038/nchem.2598

\bibitem{Ref20} Judson, R. S. \& Rabitz, H. Teaching lasers to control molecules. Physical Review Letters 68, 1500-1503 (1992). https://doi.org/10.1103/PhysRevLett.68.1500

\bibitem{Ref21} Brif, C., Chakrabarti, R. \& Rabitz, H. Control of quantum phenomena: past, present and future. New Journal of Physics 12, 075008 (2010). 

\bibitem{Ref22} Koch, C. P. \& Shapiro, M. Coherent Control of Ultracold Photoassociation. Chemical Reviews 112, 4928-4948 (2012). https://doi.org/10.1021/cr2003882

\bibitem{Ref23} Iwai, S. {\em et al.} Coherent Control of Charge and Lattice Dynamics in a Photoinduced Neutral-to-Ionic Transition of a Charge-Transfer Compound. Physical Review Letters 96, 057403 (2006). https://doi.org/10.1103/PhysRevLett.96.057403

\bibitem{Ref24} Horstmann, J. G. {\em et al.} Coherent control of a surface structural phase transition. Nature 583, 232-236 (2020). https://doi.org/10.1038/s41586-020-2440-4

\bibitem{Ref25} Paul, K. {\em et al.} Coherent control of an opsin in living brain tissue. Nature Physics 13, 1111-1116 (2017). https://doi.org/10.1038/nphys4257

\bibitem{Ref26} Zhang, Y. et al. Optical coherent quantum control of ultrafast protein electron transfer. Science Advances 11, eado9919 (2025). https://doi.org/doi:10.1126/sciadv.ado9919

\bibitem{Ref27} Duan, H.-G. et al. Intermolecular vibrations mediate ultrafast singlet fission. Science Advances 6, eabb0052 (2020). https://doi.org/doi:10.1126/sciadv.abb0052

\bibitem{Ref28} Valahu, C. H. et al. Direct observation of geometric-phase interference in dynamics around a conical intersection. Nature Chemistry 15, 1503-1508 (2023). https://doi.org/10.1038/s41557-023-01300-3

\bibitem{Ref29} Jha, A., Duan, H.-G., Tiwari, V., Thorwart, M. \& Miller, R. J. D. Origin of poor doping efficiency in solution processed organic semiconductors. Chemical Science 9, 4468-4476 (2018). https://doi.org/10.1039/C8SC00758F

\bibitem{Ref30} Matsika, S. \& Krause, P. Nonadiabatic Events and Conical Intersections. Annual Review of Physical Chemistry 62, 621-643 (2011). https://doi.org/https://doi.org/10.1146/annurev-physchem-032210-103450

\bibitem{Ref31} Boeije, Y. \& Olivucci, M. From a one-mode to a multi-mode understanding of conical intersection mediated ultrafast organic photochemical reactions. Chemical Society Reviews 52, 2643-2687 (2023). https://doi.org/10.1039/D2CS00719C

\bibitem{Ref32} Johnson, P. J. M. et al. Local vibrational coherences drive the primary photochemistry of vision. Nature Chemistry 7, 980-986 (2015). https://doi.org/10.1038/nchem.2398

\bibitem{Ref33} Polli, D. et al. Conical intersection dynamics of the primary photoisomerization event in vision. Nature 467, 440-443 (2010). https://doi.org/10.1038/nature09346

\bibitem{Ref34} Domcke, W. \& Yarkony, D. R. Role of Conical Intersections in Molecular Spectroscopy and Photoinduced Chemical Dynamics. Annual Review of Physical Chemistry 63, 325-352 (2012). https://doi.org/https://doi.org/10.1146/annurev-physchem-032210-103522

\bibitem{Ref35} Duan, H.-G., Miller, R. J. D. \& Thorwart, M. Impact of Vibrational Coherence on the Quantum Yield at a Conical Intersection. The Journal of Physical Chemistry Letters 7, 3491-3496 (2016). https://doi.org/10.1021/acs.jpclett.6b01551

\bibitem{Ref36} Duan, H.-G. \& Thorwart, M. Quantum Mechanical Wave Packet Dynamics at a Conical Intersection with Strong Vibrational Dissipation. The Journal of Physical Chemistry Letters 7, 382-386 (2016). https://doi.org/10.1021/acs.jpclett.5b02793

\bibitem {HEOM1} Tanimura, Y. Numerically "exact" approach to open quantum dynamics: The hierarchical equations of motion (HEOM). J. Chem. Phys. \textbf{153} 020901 (2020). 

\bibitem {HEOM2} Ishizaki, A. \& Tanimura, Y. Quantum Dynamics of System Strongly Coupled to Low-Temperature Colored Noise Bath: Reduced Hierarchy Equations Approach. J. Phys. Sco. Jap. \textbf{74}, 3131 (2005). 

\bibitem {Valentyn 2006} V. I. Prokhorenko, {\em et al.} Coherent Control of Retinal Isomerization in Bacteriorhodopsin. Science \textbf{313}, 1257 (2006). 

\bibitem {Shinya Koshihara coherent control} S. Koshiharaa, T. Ishikawaa, Y. Okimotoa, K. Ondab, R. Fukayac, M. Hadad, Y. Hayashie, S. Ishiharaf, T. Luty. Challenges for developing photo-induced phase transition (PIPT) systems: From classical (incoherent) to quantum (coherent) control of PIPT dynamics.  Physics Reports \textbf{942}, 1-61 (2022). 

\bibitem {Penpeng's work} P. Peng, {\em et al.} Coherent control of ultrafast extreme ultraviolet transient absorption. Nature Photonics \textbf{16}, 45 (2021). 

\bibitem{Ropers Nature CDW LEED} J. G. Horstmann, H. Böckmann, B. Wit, F. Kurtz, G. Storeck \& C. Ropers, Coherent control of a surface structural phase transition. Nature \textbf{583}, 282 (2020). 

\bibitem {Mathies_isotope} Schnedermann, C., {\em et al.} Evidence for a vibrational phase-dependent isotope effect on the photochemistry of vision. Nat. Chem. \textbf{10}, 449–455 (2018).

 
\end{thebibliography}
\end{document}